\documentclass[aps,prb,notitlepage,longbibliography,nofootinbib,reprint]{revtex4-1}
\usepackage{xcolor}
\usepackage{braket}
\usepackage{hyperref}
\usepackage{mathtools}
\usepackage{enumerate}
\usepackage{natbib}

\usepackage{setspace}
\usepackage{graphicx}
\begin{document}
\title{Long-range entanglement for spin qubits via quantum Hall edge modes}
\author{Samuel J. Elman}
\author{Stephen D. Bartlett}
\author{Andrew C. Doherty}
\affiliation{Centre for Engineered Quantum Systems, School of Physics, The University of Sydney, Sydney, NSW 2006, Australia}
\begin{abstract}
We propose and analyse a scheme for performing a long-range entangling gate for qubits encoded in electron spins trapped in semiconductor quantum dots.  Our coupling makes use of an electrostatic interaction between the state-dependent charge configurations of a singlet-triplet qubit and the edge modes of a quantum Hall droplet.  We show that distant singlet-triplet qubits can be selectively coupled, with gate times that can be much shorter than qubit dephasing times and faster than decoherence due to coupling to the edge modes. Based on parameters from recent experiments, we argue that fidelities above $99\%$ could in principle be achieved for a two-qubit entangling gate taking as little as 20 ns.
\end{abstract}
\maketitle

\section{Introduction}

Electrostatically confining electrons to quantum dots (QDs) in semiconductor heterostructures is a promising platform for the implementation of spin qubits, the fundamental building blocks of a quantum computer \cite{loss1998quantum,kloeffel2012prospects}, because the magnetic spin moments of these electrons couple weakly to the environment. Many variations of QD qubits have been proposed and demonstrated, including encoding a qubit in the spin of a single electron \cite{loss1998quantum, tarucha2006coherent, dzurak2010single, vandersypen2006driven},
the singlet-triplet qubit defined using two electrons in a double quantum dot (DQD) \cite{petta2005coherent,hrl2012}, the hybrid qubit formed of three electrons in two dots \cite{PhysRevLett.109.250503} and qubits formed of three electrons in three dots, such as exchange only qubits \cite{medford2013self}, or resonant exchange qubits \cite{medford2013quantum,Enge1500214}.

In order to perform quantum computation, scalable architectures require many qubits with high-fidelity single-qubit gates as well as high-fidelity entangling gates \cite{nielsen2010quantum}. There are many fewer experimental demonstrations of two-qubit operations for spin qubits \cite{nowack2011single,shulman2012demonstration,veldhorst2015two,nichol2016high}. An ideal method of coupling qubits would lead to two-qubit gates taking a comparable length of time and having comparable fidelity to single qubit operations. It would also allow qubits to be sufficiently widely spaced that required control and readout gates fit readily on the chip. These goals remain challenging in practice for experimental spin qubits. Designing mechanisms that can achieve high-fidelity, two-qubit entanglement for spin qubits represents the next major challenge in the realisation of a spin-based quantum computer.  
Many architectures for entangling qubits have been proposed and often investigated experimentally, including capacitive coupling \cite{taylor2005fault,stepanenko2007quantum,shulman2012demonstration,nichol2016high}, direct exchange coupling \cite{nowack2011single,veldhorst2015two}, multi-electron quantum dot mediating structures \cite{srinivasa2015tunable, braakman2013long,PhysRevB.90.045404}, electrostatic floating gate  structures \cite{trifunovic2012long,serina2016long} and their ferromagnetic equivalents \cite{trifunovic2013long}, microwave-frequency resonator couplers \cite{petersson2012circuit,viennot2015coherent,mi2016strong,PhysRevB.74.041307,russ2015long,srinivasa2016entangling}, photon assisted coupling \cite{stano2015fast}, and phonon assisted coupling \cite{schuetz2015universal}. 

There have been earlier proposals for coupling qubits using quantum Hall edges. One approach is to have sufficient tunnel coupling to create an excitation of the edge that travels between the two qubits\cite{stace2004mesoscopic,benito2016dissipative}. Recently, it has been suggested that the spin degree of freedom of conducting edge states of quantum Hall liquids can be used to entangle spin qubits over long distances, making use of the magnetic interactions, and mediated by the conducting edge states of quantum Hall (QH) liquids, to which the QDs are tunnel coupled \cite{yang2016long}. The advantage of using QH edge states is twofold: firstly, the edge states and the QDs can be formed in the same material (by top gates, or etching) in a two-dimensional electron gas (2DEG). Secondly, the QH edge states are much more robust against disorder effects than quasi-one-dimensional conduction channels, which might be used in the same way\cite{stace2004mesoscopic}. 

In this paper, we propose an alternative mechanism to achieve long-distance entanglement of spin qubits using an interaction between confined electron spins in QDs, mediated by the conducting edge states of a QH droplet to which the qubits are \emph{electrostatically} coupled. Since our proposed system of two-qubit entanglement via QH edge modes makes use of an electrostatic interaction rather than the magnetic interaction in Reference~\onlinecite{yang2016long}, our system requires that the electron tunnelling into and out of the edge modes be prohibited.  Our proposed two-qubit entangling gate is based on a coupling of the electric dipole of the qubit, which is state-dependent, with the edge modes of the QH droplet described as a quantum harmonic oscillator\cite{mahoney2016chip,cano2013microwave}.  We obtain a qubit-state-dependent force on the oscillator, resulting in a general form of coupling that has been used for entangling gates in a variety of other physical systems, including trapped ions~\cite{Wineland2003,milburn2000ion,sorensen1999ion} and longitudinally-coupled circuit QED qubits~\cite{nakamura2015,royer2016fast,divincenzo2016}.  We demonstrate that this mechanism can lead to strong coupling with low decoherence, and as such is a promising candidate for two-qubit entanglement.  In particular, using parameters from recent experiments for singlet-triplet qubits \cite{nichol2016high} as well as the exploration of edge modes in QH droplets\cite{mahoney2016chip}, we predict that the effective qubit-qubit coupling can be as high as 60 MHz, leading to entangling gate times of order 20 ns and resulting gate fidelities greater than 99\%. One advantage of the scheme is that it can couple qubits over lengths of order tens of micrometres. This alleviates the crowding that would result from attempting to couple quantum dot qubits using direct exchange coupling. The scheme is similar to a recent proposal for coupling spin qubits using an oscillator\cite{schuetz2017high}, in this work we envisage modulating the coupling such that the gate operates with high fidelity even with a large qubit-state energy splitting.

The electrostatic interaction requires that we use a qubit implementation whose spin state can be mapped onto a charge state.  For concreteness, we focus on singlet-triplet qubits formed in GaAs-AlGaAs heterostructures, where the QH droplet can also be formed in the 2DEG.  With singlet-triplet qubits, both charge and spin degrees of freedom of the electrons play a role \cite{petta2005coherent}.   We emphasise, however, that the general approach taken in this paper can be adapted to other breeds of qubit, such as hybrid and resonant exchange qubits.

A singlet-triplet qubit configuration is advantageous for several reasons: firstly, due to the presence of two electron spins they are robust to background noise and so exhibit longer dephasing times \cite{nichol2016high}. Secondly, state preparation and manipulation can be achieved predominantly using electric fields, rather than magnetic fields which are slower to vary or switch on and off. Thirdly, the spin states of the qubits can be mapped to charge states, which results in straightforward state readout \cite{petta2005coherent,elzerman2004}.

Our paper is structured as follows.  In Sec. \ref{sec:qubit_qubit_coupling} we derive the expression for the qubit-qubit entanglement from the general Hamiltonian. Then we derive an expression for the strength of the coupling by investigating the qubit-edge electrostatic coupling, and calculate the qubit-edge coupling for some realistic parameters in Sec. \ref{sec:electrostatic_coupling}. Finally, in Sec. \ref{sec:implement_gate}, we analyse the two-qubit gate, using the average gate fidelity as a metric.

\section{Qubit-Qubit Coupling}\label{sec:qubit_qubit_coupling}

In this section, we describe how a two-qubit entangling gate can be achieved by coupling the qubits individually to the edge modes of a QH liquid, using a theoretical framework that has predominantly been used for superconducting qubits and quantum optics\cite{gambetta2008quantum}, but which easily applies to this system.

\begin{figure}[b]
\includegraphics[width=0.9\linewidth]{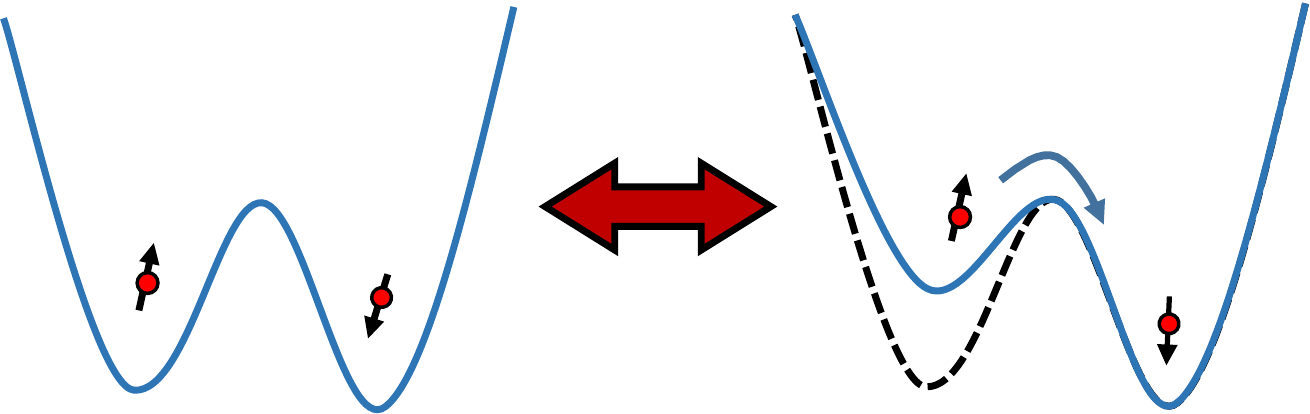}

\caption{Diagrammatic representation of driven oscillations of the state-dependent electric dipole moment. A gate voltage defining the quantum dots can bias the potential wells in such a way as to favour tunnelling of the left spin into the right dot.  Pauli spin blockade prevents this tunnelling if the spins are in a triplet state.   If this gate voltage is modulated, the result is an oscillating qubit-state-dependent force on the edge modes of the nearby quantum Hall droplet. \label{fig:oscillations}}
\end{figure}

The essential mechanism for coupling is as follows.
The singlet-triplet spin qubits can be brought to an operating point where they possess a state-dependent electric dipole moment due to the Pauli spin blockade, as experimentally demonstrated in Ref.~\onlinecite{petta2005coherent}.  By driving oscillations in this state-dependent electric dipole moment, as depicted in Fig.~\ref{fig:oscillations}, we can excite the edge modes of a nearby QH droplet.  The driving of a quantum harmonic oscillator (the QH edge modes) by a qubit-state-dependent force is a coupling mechanism that has been well studied in a variety of qubit architectures, and with multiple qubits coupled to the oscillator, it can be used to generate a two-qubit entangling gate, e.g., as in trapped ions~\cite{Wineland2003,milburn2000ion,sorensen1999ion} and longitudinally-coupled circuit QED~\cite{nakamura2015,royer2016fast,divincenzo2016}.

For this entangling gate to work with high fidelity, the qubit-oscillator coupling must exceed the decay rate of the oscillator.  Intuitively, if an excitation of the oscillator occurs during the coupling and this excitation is subsequently lost from the cavity, then the gate does not succeed.  This decoherence mechanism can be minimised by ensuring that the oscillator does not become excited with even a single excitation during the coupling process; this can be enforced by demanding that the detuning of the drive frequency from the cavity frequency is large.

\begin{figure}
\includegraphics[width=0.9\linewidth]{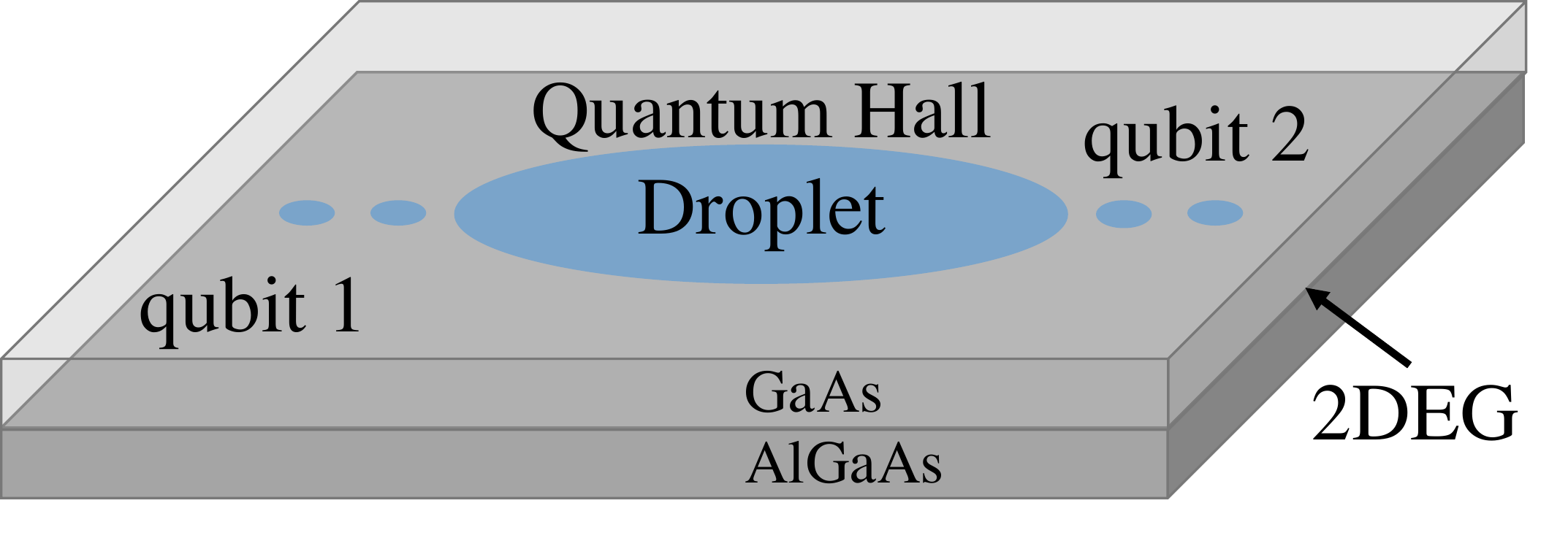}
\includegraphics[width=0.9\linewidth]{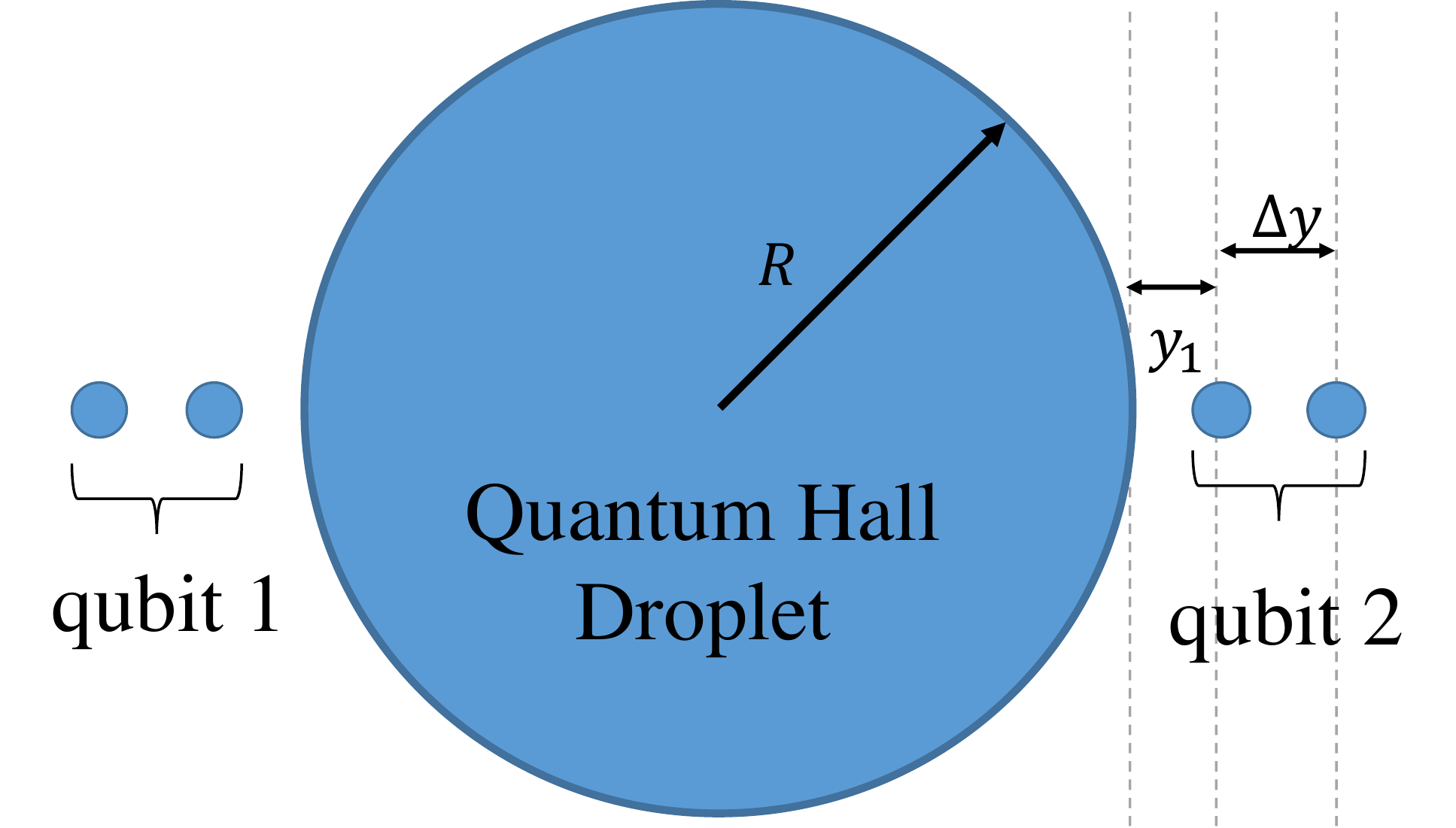}

\caption{Schematic diagram of the system for qubit-qubit coupling: a disc of radius $R$ is formed from a 2DEG engineered in a GaAs/AlGaAs heterostructure. A singlet-triplet qubit, of inter-dot separation $\Delta y$, is formed radially  on either side. The qubits are electrostatically coupled to the edge modes of the QH disc. By driving oscillations in the qubits, we can effectively couple multiple qubits to one another using the QH disc as a mediator.\label{fig:schematic}}
\end{figure}

Because the qubit-qubit entanglement mediated by the QH edge modes is dependent on the qubits being driven at a particular frequency, the entangling gate can be switched on (off) by gradually increasing (decreasing) the amplitude of the driven oscillations. This is a control feature that `static' coupling schemes (such as the floating gate `dog-bone' structures \cite{trifunovic2012long,trifunovic2013long,serina2016long}) do not possess. In principle, this allows multiple qubits to be placed around a single QH disc, which can then be coupled selectively.

The basic configuration for the qubit-qubit coupling is shown in Fig~\ref{fig:schematic}. A circular QH droplet of radius $R$ is formed out of two dimensional electron gas (2DEG).
Multiple double quantum dot (DQD) qubits, of inter-dot separation $\Delta y$, are defined radially outside of the edge of the QH disc a distance of $y_1$ from the edge. Note that the angle of separation between the qubits at the edge does not affect the coupling.

The QH disc is formed by depleting electrons in the 2DEG to form a finite region, which is then subjected to a large magnetic field. It is well known that this produces propagating edge modes around the circumference of the disc, known as edge magneto plasmons (EMPs)\cite{wen2004quantum,fradkin2013field}. The fundamental mode of the EMPs (with frequency $\omega_c$) can described as a quantum harmonic oscillator, in terms of annihilation (creation) operators $a(a^\dagger)$. 
Two qubits are formed radially, adjacent to the edge of the QH disc, each electrostatically coupled to the edge mode as shown in Fig~\ref{fig:schematic}. The qubits are driven at a frequency $\omega_{di}$, which allows them to be coupled to the edge, with coupling strength $g_i$, an expression, which will be derived in Sec.~\ref{sec:electrostatic_coupling}. Taking into account the driving of these qubits, we can use a model that is similar to that presented in Ref.~\onlinecite{royer2016fast} for superconducting qubits. Our analysis of the model also bears similarities to Ref.~\onlinecite{royer2016fast}, however we make slightly different approximations in order to model the operation of the gate in the regime that we anticipate will be most relevant for future experiments in this system. In particular we will perform the two-qubit gate in a regime in which the oscillator never becomes significantly excited. 

In our model, the system is described by the Hamiltonian:
\begin{multline}
H=\,\omega_\text{c} a^{\dagger}a+J_1\sigma_{z1}/2+J_2\sigma_{z2}/2 \\
+\big[g_1\cos(\omega_{d1}t)\sigma_{z1} +g_2\cos(\omega_{d2}t)\sigma_{z2}\big](a+a^{\dagger})
\label{eq:ham_full}
\end{multline}
where we have set $\hbar=1$. The first term in Eq.~\ref{eq:ham_full} describes the oscillator (the edge mode of the QH droplet), the second and third terms describe the qubits, and the final term describes the coupling of the qubits to the edge mode.  This coupling term captures the essential mechanism described above.  We note that the microwave drive on the bias of the qubit that is required to obtain the time dependence of this equation will also result in a direct drive of the qubit.  We have omitted this qubit term because this drive will be very far off resonance.  Note that the qubits are assumed to be driven in phase with each other.

We drive the qubits at the same frequency, $\omega_d$, and shift to a rotating frame at this frequency, that is, we move into a frame where the free evolution is such that the qubits are stationary and the edge mode oscillates at a detuned frequency of $\Delta=\omega_c-\omega_d$. This gives us the interaction Hamiltonian in the rotating frame with respect to $\omega_d$:
\begin{multline}
H_\text{int}=\Delta a^\dagger a+g_1(a+a^\dagger)\sigma_{z1}/2 \\ +g_2(a+a^\dagger)\sigma_{z2}/2.
\label{eq:ham_int}
\end{multline}
where we have also used the rotating wave approximation to remove rapidly oscillating terms.

In order to capture the full system dynamics, including the damping of the edge modes and qubit decoherence, we model the system using a master equation \cite{wiseman2010quantum}. Our modelling of the qubits focusses on charge noise, since this is the dominant noise process for most recent experiments with singlet-triplet qubits\cite{nichol2016high,shulman2012demonstration,dial2013charge}. We note that the decoherence due to noise in the Overhauser field can also be added to this model in a straightforward manner. In order to match experimental observations more closely, our model involves two components of charge noise\cite{wardrop2013two}. A high frequency component results in exponential decay and is modelled by Lindblad terms in the master equation. The strength of this noise can be determined by the single qubit dephasing time, $T_2$. A low frequency component of charge noise is modelled by averaging over a Gaussian distribution of the qubit exchange splittings $J_1$ and $J_2$. The width of the distribution can be determined by the single qubit ensemble dephasing times, $T_2^*$ \cite{dial2013charge}. The QH edge mode is damped, and the resulting exponential decay can readily be modelled by a master equation. This decay rate $\kappa=\omega_c/2Q$ will depend on the device but reasonable values can also be obtained by comparing to recent measurements\cite{mahoney2016chip}.

In order to demonstrate the two-qubit interaction in our system we initially consider only the master equation that deals with high frequency charge noise on the qubits and edge mode damping. The appropriate master equation is:
\begin{equation}
\dot\rho=-i[H_\text{int},\rho]+2\kappa\mathcal{D}[a]\rho+\sum_{i=1}^2\gamma_{\phi i}\mathcal{D}[\sigma_{zi}]\rho/2
\label{eq:master}
\end{equation}
where $\gamma_{\phi i}=1/T_{2i}$ is the dephasing rate of the $i$-th qubit, and $\mathcal{D}[c]$ represents the usual dissipation super-operator $\mathcal{D}[c]\rho\equiv c\rho c^\dagger-c^\dagger c\rho/2-\rho c^\dagger c/2$ (see Ref.~\onlinecite{wiseman2010quantum}). 

The edge mode's mediation of a two-qubit interaction can be made explicit by applying a polaron transformation\cite{gambetta2008quantum}. The polaron transformation is a qubit-state-dependent displacement of the edge mode oscillator and is defined as the unitary operator: ${U}=\exp{[\sum_{i=1,2}(\alpha_i\sigma_{zi}a^\dagger-\alpha_i^*a\sigma_{zi})]}$. The displacements $\alpha_i$ are chosen so that terms linear in $a$ and $a^\dagger$ cancel out after the polaron transformation is applied to the master equation.

Performing the polaron transformation, allowing $\alpha_i=g_i/2(\Delta+i\kappa)$, and assuming $\kappa\ll\Delta$, takes the master equation of Eq.~\eqref{eq:master} to the following ``polaron-shifted" master equation, where we denote the polaron-shifted operators, such as $\tilde{\sigma}_{z1}=U\sigma_{z1}U^\dagger$, with tildes:
\begin{multline}
\dot{\rho}=-i[H_\text{pol}, \rho]+2\kappa\mathcal{D}[\tilde a]\rho+\gamma_{\phi1}\mathcal{D}[\tilde \sigma_{z1}]\rho/2\\
+\gamma_{\phi2}\mathcal{D}[\tilde \sigma_{z2}]\rho/2+\Gamma_d\mathcal{D}[(\tilde \sigma_{z1}+\tilde \sigma_{z2})]\rho/2
\\-\frac{\kappa}{\Delta-i\kappa}[\rho(g_1\tilde \sigma_{z1}+g_2\tilde \sigma_{z2}),\tilde a]\\-\frac{\kappa}{\Delta+i\kappa}[\tilde a^\dagger,\rho(g_1\tilde \sigma_{z1}+g_2\tilde \sigma_{z2})]
\,.
\label{eq:master_pol}
\end{multline}
$\Gamma_d=\kappa g_1g_2/2(\Delta^2+\kappa^2)$ is the rate of correlated dephasing of the qubits that is associated with the relaxation of the QH oscillator. $H_\text{pol}$ is the new, polaron-shifted Hamiltonian, which has an explicit two-qubit coupling:
\begin{equation}
{H}_\text{pol}=\Delta {a}^\dagger{a}+J_{12}
\tilde \sigma_{z1}\tilde \sigma_{z2} \,,
\label{eq:ham_pol}
\end{equation}
where:
\begin{equation}
J_{12}(t)=-\frac{g_1(t)g_2(t)}{2\Delta}\frac{\Delta^2}{\Delta^2+\kappa^2}\,.
\label{eq:q-q_coup}
\end{equation}
In the polaron shifted picture there is no longer a Hamiltonian coupling of the oscillator to the qubits. This interaction is replaced by a direct coupling of the two qubits. This Ising-type coupling is well known to lead directly to non-trivial two-qubit gates such as the controlled phase (cPHASE) gate.

In this polaron shifted picture, it is important to remember that the Pauli spin matrices $\tilde \sigma_{zi}$ no longer correspond to the bare physical qubits described in Eqs.~\eqref{eq:ham_full} and \eqref{eq:ham_int}, but rather qubits that are dressed by excitations of the QH oscillator. Likewise the effective oscillator mode described by $\tilde{a}$ is dressed by the presence of the qubits and the qubit drives. Inspecting the polaron transformation $U$ with the chosen values for $\alpha_i$ makes it clear that the distinction between the bare and dressed qubits becomes less significant for large $\Delta\gg g$. Considerable physical insight can be gained by studying the behaviour of the dressed qubits. For example, in Sec.~\ref{sec:implement_gate} we will compute approximate expressions for the fidelity of the gate performed on the dressed qubits. This can then be regarded as an estimate of the true fidelity of the gate performed on the physical qubits, with corrections anticipated to be of higher order in $g/\Delta$. These simple estimates can then be tested by comparison to simulations of the full master equation~(\ref{eq:master}).

The polaron-shifted master equation~(\ref{eq:master_pol}) involves much weaker coupling of the qubits and the QH oscillator. Moreover, it is easy to see that the dressed quantum Hall oscillator relaxes to its vacuum state and remains there at long times. Consequently, we can obtain an approximate master equation for the dressed qubits alone by using the ansatz $\rho\simeq \rho_q\otimes |0\rangle\langle 0|$ and then tracing out the QH oscillator. We obtain:
\begin{equation}\begin{split}
\dot\rho_q=-i[J_{12}\tilde{\sigma}_{z1}\tilde{\sigma}_{z2},\rho_q]+\gamma_{\phi1}\mathcal{D}[\tilde \sigma_{z1}]\rho_q/2+\gamma_{\phi2}\mathcal{D}[\tilde \sigma_{z2}]\rho_q/2\\+\Gamma_d\mathcal{D}[(\tilde \sigma_{z1}+\tilde \sigma_{z2})]\rho_q/2.
\label{eq:master_high}
\end{split}
\end{equation}
Here the first term describes the ideal qubit evolution under the polaron shifted Hamiltonian of Eq.~\eqref{eq:ham_pol}, the second and third terms describe the single qubit dephasing due to high frequency noise and the final term describes the dephasing of the qubits due to the presence of the lossy QH oscillator. In subsequent sections we supress the subscript on $\rho_q$ since it will be clear from the context whether a qubit density matrix, or a density matrix for the full system, is intended.

In the limit that there is no noise in the system, modulating the qubit-qubit coupling (proportional to $\sigma_{z1}\sigma_{z2}$) for a time $t_g = \pi/(2|J_{12}|)$, the evolution under Eq.~\eqref{eq:ham_pol} is equivalent to the entangling cPHASE gate $U_{CP} (\pi) = \text{diag}(1, 1, 1, e^{i\pi})$ up to single qubit $Z$ rotations, with entangling gate time given by
\begin{equation}
t_g=\frac{\pi\Delta}{g_1g_2}\frac{\Delta^2+\kappa^2}{\Delta^2}.
\label{eq:gate_time}
\end{equation}
When noise is included, modulation of the qubit-qubit coupling for $t_g$ yields an approximate cPHASE gate, and its performance can be quantified by the gate fidelity.  We will analyse this gate fidelity in Sec.~\ref{sec:implement_gate} for multiple noise regimes, using both the polaron picture master equation of Eq.~\eqref{eq:master_high} as well as the full master equation of Eq.~\eqref{eq:master}.

\section{Electrostatic qubit-edge mode coupling}\label{sec:electrostatic_coupling}

In this section, we will model the electrostatic coupling between the qubit and the edge mode of the QH droplet, $g$, and derive an expression for the magnitude of this coupling as a function of the parameters of the geometry of the system: the radius of the disc $R$, the qubit edge separation $y_1$, and the interdot separation $\Delta y$. We then estimate the magnitude of this coupling using values for these parameters consistent with current experiments. As discussed, there is a state-dependent electric dipole associated with the qubit, which we want to evaluate in order to give an estimate for the coupling between the qubit and the oscillator modes.  In our analysis the electric dipole of the qubit is assumed to be oriented perpendicular to the edge of the disc, which allows for maximum coupling to the edge modes.

Following Ref.~\onlinecite{taylor2005fault}, the qubits are described using logical basis states: $\ket{S}=\sin\theta_q\ket{(0,2)S}+\cos\theta_q\ket{(1,1)S}$ and $\ket{T}=\ket{T(1,1)}$. Here, $\theta_q=[0,\pi/2)$ is a parameter that describes the extent to which the singlet's wave function is weighted towards the $(0,2)$ charge distribution. Thus the two logical spin basis states are represented by two distinct charge distributions: the triplet distribution being symmetric and the singlet distribution biased to one side. 
Again, we emphasise that in our coupling scheme we wish to work in a regime where both tunnel coupling and exchange coupling to the edge are negligible, and only consider the electrostatic coupling of the qubit to the edge.  This scheme differs from Ref.~\onlinecite{yang2016long}, where the main process of coupling qubits to the edge is through the exchange interaction. 


The electrostatic coupling strength $g$ between the qubit and the edge modes is calculated by the change in energy of the edge mode when the qubit changes from the $\ket{S}$ state to the $\ket{T}$ state. To do this calculation, we first examine the case when a single electron has been moved completely from one QD to the other. We then account for only a portion of the singlet wave function shifting into the $(0,2)$ charge state by introducing a multiplicative factor of $\sin^2\theta_q$.
 Within this multiplicative factor of $\sin^2\theta_q$ is an implicit dependence on the an electric field bias placed on the qubits.  In our protocol, we will propose to modulate this electric field bias at a frequency $\omega_d$ near the edge-mode resonant frequency.  Consequently, the parameters $g_1$ and $g_2$ will be proportional to the amplitude of this microwave drive.  However, this driving must remain in a regime of sufficiently small $\theta_q$ such as to avoid Landau-Zener transitions to higher energy states.  This multiplicative factor also includes an assumption that the oscillating field is driven off resonance to the QH droplet, meaning that there is minimum direct coupling between this field and the droplet.

We make the approximation that the electrons in the qubit are point like, and reside at the centre of the QDs (an approximation verified in Ref.~\onlinecite{serina2016long}).  We also approximate the edge modes as one-dimensional, because the width of the edge is smaller than magnetic length $\ell_B$, which sets the relevant length scale \cite{wen2004quantum}.
We also include a factor $\eta$ that accounts for the screening of electric fields by the metallic gates that must surround the system in order to form the DQD and QH disc.
{We estimate the screening factor using the electron microscope images from Ref.~\onlinecite{shulman2012demonstration}, by taking the ratio of the electric field that ends on metallic gates and the electric field that ends on the edge of the disc; we estimate that approximately 40\% of the electron's electric field ends on the metallic gates, and so the factor for the electrostatic screening is taken to be $\eta\simeq0.60$}. This is consistent with previous theoretical descriptions of the effect of the metallic gates on the electric field density and scalar potential \cite{stopa1996quantum}.

There are two contributions to the electrostatic coupling: (i) the driving of the edge modes due to the shifting potential energy due to the $i$-th qubit proximal to the edge, $g_i$, which appears explicitly in Eq.~\eqref{eq:ham_full}; (ii)  {the shift in the edge mode's frequency due to the state of the qubit  modulating the velocity of the EMP, $g_\text{vel}=\hbar\Delta\omega_c$}. The two contributions are independent of one another; however, as we will show,  contribution (i) is orders of magnitude larger than (ii).

Considering contribution (i):  the electric potential energy of the QH edge due to the electric potential of the electrons in the quantum dot qubits is $U=\int \rho(s)V(s)ds$, where $\rho(s)$ is the linear charge density along the QH edge, and $V(s)$ is the electric scalar potential. The co-ordinate $s$ parameterizes the edge of the QH droplet. The potential around the edge of a disc of radius $R$, due to a point charge placed $r$ away from the edge of the disc (that is, $r=R+y$ away from the centre of the disc) is:
\begin{equation}
V(\theta,R+y)=
\frac{{e}/{4\pi\varepsilon}}{\sqrt{((R+y)^2+R^2-2R(R+y)\cos\theta)}}
\label{eq:potential}
\end{equation}
where $\varepsilon=\varepsilon_0\varepsilon_r$ is the permittivity of the material in which the 2DEG is formed, and $\theta$ is the angle around the disc.
Equation \eqref{eq:potential} describes how, as the electron is moved closer, the electrostatic potential increases. This change in the potential modulates the energy of the EMP that travels around the disc. This change in potential energy is $g_\text{pot}$.

In our analysis the disc is assumed to be perfectly circular, for simplicity.  However, we emphasise that similar results will hold for a QH droplet of any shape, with Equation \eqref{eq:potential} modified accordingly for the potential along any closed loop in the presence of a point charge.  We note that the coupling between the qubits and the edge modes will be largest when the curvature of the QH droplet at the point nearest the qubits is minimized.  There is an opportunity, then, to consider improving this coupling by using non-circular droplets; however, we leave such an analysis for future work.

The EMP modes of the quantum Hall edge form standing wave modes $a_n$ having wavelength $2\pi R/n $ and frequency $nv/2\pi R$, where $v$ is the velocity of the EMPs.  We can express the charge density $\rho(s)$ around the edge in terms of the angle $\theta =  s/R$ and QH filling factor $\nu$ as\cite{wen2004quantum}:
\begin{equation}
\rho(\theta)=e\sum_n\frac{\sqrt{\nu n}}{(2\pi R)}e^{in\theta}(a_n+a_n^\dagger).
\end{equation}
The integral for the potential energy at the edge due to a point charge at a distance $y$ outside the radius of the disc is given by:
\begin{equation}\begin{split}
U_{\nu,n}(y)=&\frac{\sqrt{\nu n}}{L}\frac{e^2}{4\pi\varepsilon}\\&\times \int_{-\pi}^\pi\frac{e^{in\theta}Rd\theta}{\sqrt{(R+y)^2+R^2-2R(R+y)\cos\theta}},
\label{eq:potential_int}
\end{split}
\end{equation}
where $L=2\pi R$ is the circumference of the disc. This integral can be solved to obtain an expression in terms of complete elliptical integrals \cite{abramowitz1964handbook}. The electrostatic coupling of a single qubit to the EMP is then $g_\text{pot}=U_{\nu,1}(y_1)-U_{\nu,1}(y_1+\Delta y)$. 

\begin{figure}
	\centering
	\includegraphics[width=0.95\linewidth]{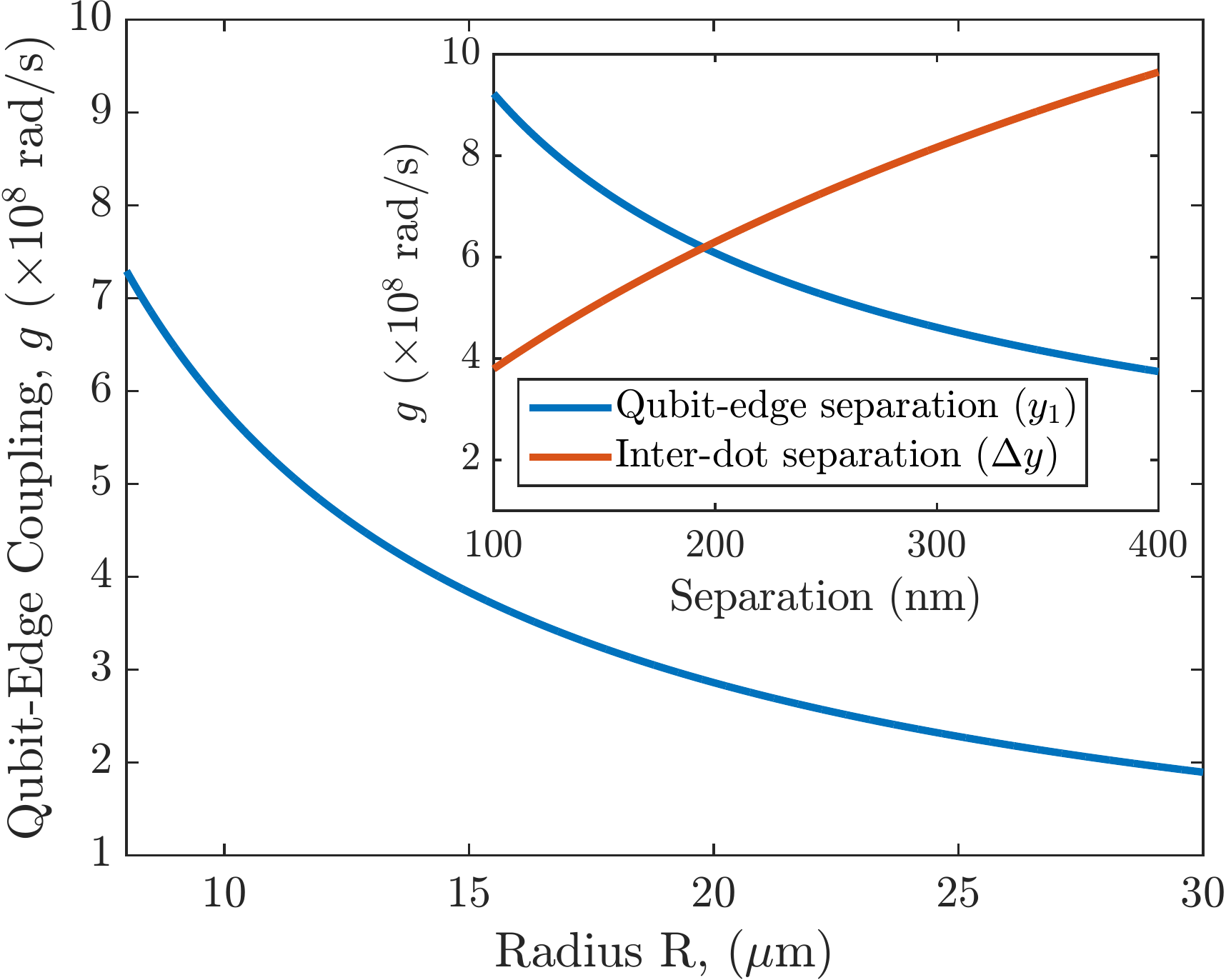}
	\caption{Qubit-edge coupling given by Equation \eqref{eq:coupling_t} as a function of disc radius, $R$, with qubit-edge separation and inter-dot separation fixed at $y_1=150$nm and $\Delta y=250$nm respectively; (inset) qubit-edge separation, $y_1$ (with disc radius and interdot separation fixed at $R=8\mu$m and $\Delta y=250$nm, respectively) and inter-dot separation (for $R=8\mu$m and $y_1=150$nm).}
	\label{fig:g_t_para}
\end{figure}

It is perhaps more illustrative to consider the expansion of these elliptic integrals in terms of elementary functions. Under the assumption that the radius of the disc is much larger than the separation of either DQD ($R\gg y_i$), we obtain the following expression for the coupling of a qubit to the first harmonic mode of the EMP ($n=1$):
\begin{equation}\begin{split}
g_i\simeq\eta\sin^2\theta_q\frac{\sqrt{\nu}}{L}\frac{e^2}{2\pi\varepsilon}\Bigg\{\ln\left(\frac{y_1+\Delta y}{y_1}\right)\left[1-\frac{y_1}{2R}
\right]\\+\frac{\Delta y}{2R}\left[2-2\ln2-\ln\left({\frac{y_1+\Delta y}{2R}}\right)\right]\Bigg\}.
\label{eq:coupling_t}
\end{split}\end{equation}
The qubit-edge mode coupling depends on the separations $y_1$ and $\Delta y$, and the radius of the disc, $R$, as shown in Fig.~\ref{fig:schematic}, as well as the physical parameters $\eta$ and $\theta_q$. We now discuss the dependence of the qubit-edge coupling with these parameters, as illustrated by Equation~\eqref{eq:coupling_t}, and Fig.~\ref{fig:g_t_para}.  In order to maximise $g$, the lengths $R$ and $y_1$ must be made as small as is experimentally feasible, while $\Delta y$ must be made as large as possible. Reducing $R$ means that more of the edge of the QH disc will fall within a region of higher electrostatic potential on the edge of the disc. Similarly, reducing $y_1$ results in the edge resting within the region of higher electric potential at the edge of the QH disc. Increasing $\Delta y$ means that there is a greater difference in the potential at the edge as an electron is moved from one QD to the other, so that there is a more measurable difference in the electric potential at the edge. However, engineering and practical limitations exist on all three, meaning $R$ and $y_1$ cannot be made arbitrarily small and $\Delta y$ cannot be made infinitely large. 

We now consider these limitations based on recent experiments:

\textit{Limitations on $R$}---As the disc is reduced in size ($R\rightarrow0$), the edge modes couple to bulk magnetoplasmons and this leads to a breakdown in the assumption that the edge of the disc is one-dimensional \cite{fradkin2013field}. If the edge magnetoplasmons couple to those of the bulk, it is no longer accurate to discuss the edges modes of the QH disc at all, and the qubit may become coupled to the entire disc, which is much more susceptible to noise, reducing the effectiveness of the QH disc as a mediator for qubit-qubit coupling.
There are also experimental considerations: as the disc is reduced in size, the frequency of the edge modes increases. An increase in frequency results in stronger coupling, however, there is a practical limit to the frequency before the system moves into the microwave regime which is expected to lead to reduced lifetimes of the edge modes, as well as other experimental complications. To avoid entering this regime, we limit the maximum frequency of the EMP modes to approximately $f_\text{EMP}\sim 8$ GHz. Given a conservative estimate of the velocity of the edge modes of $v\sim4\times10^5$ m/s (see Refs.~\onlinecite{mahoney2016chip, kumada2011edge}), this leads to a minimum radius of our disc at $R_\text{min}=8\,\mu$m. Thus we will use a fundamental frequency of $f=8$ GHz, and a radius of $R=8\mu$m.

\textit{Limitations on $y_1$}---If the QD is formed too close to the edge of the QH disc (if $y_1$ is too small) electrons in the qubit could tunnel into the QH disc. From recent experiments  we estimate this limit to be approximately $y_1\simeq100$ nm, as barriers as small as this have been shown to be effective \cite{kataoka2009coherent}.  For our analysis, we will use a value of $y_1=150$ nm, which is consistent with recent experiments that create tunnel barriers over hundreds of nanometers \cite{elzerman2004}. 

\textit{Limitations on $\Delta y$}---As the inter-dot separation increases ($\Delta y\rightarrow\infty$) the two electrons in the DQD can become uncoupled. Two electron coupling, to form singlet and triplet states, has been achieved in quantum dot structures over distances as high as $\Delta y_\text{max}\simeq400$ nm \cite{giavaras2007singlet}, however, this is much larger than the inter-dot separation of $\Delta y\simeq200$ nm, which is regularly used for singlet-triplet qubits. For this reason, in our analysis we consider the inter-dot separation to be $\Delta y=250$ nm, which is larger than normal, but still consistent with recent experiments \cite{PhysRevLett.105.246804}.

We also note that the strength and direction of the external magnetic field will be relevant for both the QH droplet and the spin qubits.  To enter the quantum Hall regime, we require an external magnetic field perpendicular to the 2DEG with strength of order 1~T.  Double quantum dot spin qubits have typically used external magnetic fields parallel to the 2DEG, of magnitude of order 100~mT, to provide the Zeeman splitting of unwanted spin states.  Magnetic field components that are perpendicular to the 2DEG can have an effect on the orbital states of quantum dots, especially for large dots.  For singly- and doubly-occupied quantum dots as studied here, the spectral properties are essentially unchanged for perpendicular magnetic fields up to about 1~T (Ref.~\onlinecite{Zumbuhl}).  If larger perpendicular fields are required, one may instead use inverted singlet-triplet qubits as described in Ref.~\onlinecite{Mehl2014}.

This magnetic field is required to achieve a filling factor of $\nu=2$, where we have assumed the 2DEG density is similar to that of Ref.~\onlinecite{mahoney2016chip}. Given that the magnitude of the magnetic field is dependent on the density of the 2DEG, it is reasonable to consider how lower-density samples could allow the required magnetic field strength to be reduced. 

A magnetic field gradient is also required for full qubit control. In current experiments, magnetic field gradients are created in two different ways: either from a micro-magnet\cite{Tarucha} or from dynamic nuclear polarisation\cite{nichol2016high}.   As the latter approach places stringent requirements on the external magnetic field and its orientation, it may be preferable to use a micro-magnet in this case.

For our analysis, we assume a 2DEG in a GaAs heterostructure. Half of the electron's electric field will exist outside of the semiconductor slab, as a result we take the average of the permittivities of GaAs and air (or vacuum).  We use $\varepsilon_r\simeq7$ in our analysis\cite{XX}. With a QH disc of radius $R=R_\text{min}=8\,\mu$m, and QD-edge separation of $y_1=150$ nm and inter-dot separation $\Delta y=250$ nm, we calculate a qubit-edge coupling of $g_i/\hbar\simeq7.29\times10^8$ rad/s, or alternatively $g_i/h\simeq58.1$ MHz.  Here we will take $\theta_q\simeq0.1\pi$ based on estimates made in Reference~\onlinecite{taylor2005fault}. 
The estimated qubit-edge coupling is large, and this strength gives cause for optimism.

Now considering contribution (ii): the velocity of the EMP is proportional to the perpendicular electric field at the edge\cite{wen2004quantum}, i.e., $v(s)=E_\perp(s)/B$. The electric dipole of the QD will then increase the perpendicular electric field along the edge over a short distance and therefore speed up the EMP.
{The electric field of a point charge follows an inverse square law ($E\propto1/r^2$), which means at large distances from the electron, where the curvature of the edge of the disc could be detected, the contributions of electric field will be negligible. Thus, for the electric field, the edge can be approximated by an infinite straight line.}

The change in energy $g_\text{vel}$ due to the velocity modulation is calculated using the relationship for the change in edge mode frequency ($\Delta \omega$) in terms of the change of EMP velocity ($\Delta v(s)$) due to the position of the electron. Using the expression for the velocity in terms of electric and magnetic fields, and then expanding to first order, gives:
\begin{equation}
\begin{split}
g_\text{vel}=&\eta\sin^2\theta_q \\&\lim_{\ell\rightarrow\infty}\int_{-\ell/2}^{\ell/2}\frac{(E_\perp(s,y_1)-E_\perp(s,y_1+\Delta y))}{2\pi R^2B}ds
\end{split}
\end{equation}
where the electric field along the edge, due to an electron at distance $y$ away from the edge, is given by $E_\perp(s,y)={ey}/(4\pi\varepsilon(x^2+y^2)^{3/2})$.  We calculate the contribution to the coupling from the shift in the EMP's frequency using the same geometric parameters as described for contribution (i), with a magnetic field of $B\simeq1T$, to be $g_\text{vel}/\hbar\simeq1.7\times10^6$ rads$^{-1}$. Thus, for radii much larger than qubit edge and inter-dot separations, $R\gg (y_1, y_1+\Delta y)$,  contribution (ii) is negligible in comparison to contribution (i), and perhaps not even visible above background charge noise \cite{dial2013charge}, so can be neglected in further analysis. In the next section, we use the value calculated for qubit-edge coupling due to the shift in the potential energy, $g_i$, to analyse the two-qubit entangling gate of Eq.~\eqref{eq:ham_pol}.

\section{Implementation of a two-qubit gate}\label{sec:implement_gate}

In this section, we analyse the quality of the two-qubit entangling gate defined in Sec.~\ref{sec:qubit_qubit_coupling} for a variety of parameters, in particular the detuning and qubit edge coupling. 
The figure of merit with which we will assess a logical gate is the average gate fidelity, $\bar{F}$.
The average fidelity is the chosen figure of merit because of its wide use throughout the literature, and because it can be well approximated experimentally using randomised benchmarking. We use this section to demonstrate how the average fidelity of the two-qubit gate depends on parameters such as the detuning, $\Delta$, and qubit-edge coupling, $g$. Ultimately, we show that there is an optimal detuning for which the average fidelity is maximized, find a lower bound on the qubit-edge coupling---as we show, $g$ must be at least an order of magnitude greater than the edge decay rate---and give an estimate for the average fidelity of the gate.

We investigate the effect of high frequency charge noise on the qubits using numerical simulations of both the bare qubits using the unshifted master equation~(\ref{eq:master}) and the dressed qubits using Eq.~(\ref{eq:master_pol}). This will allow us to compare the true fidelity of the gate applied to the bare qubits to the fidelity of the gate performed on the dressed qubits. We will find that we can both improve the fidelity of the physical gate and increase the agreement between the physical and dressed fidelities by using smoother gate pulses that prevent polaron oscillations in the fidelity. We carry out a detailed investigation of the optimal pulse-shape and timing. We then consider the effect of low frequency charge noise and derive analytic approximations to the gate fidelity from Eq.~(\ref{eq:master_high}) taking into account both sources of charge noise. We show that these approximations are accurate when compared to full numerical simulations. The approximations allow us to analyse the optimal choice of detuning $\Delta$ and finally, we find an estimate of the optimal gate fidelity that can be achieved using physically reasonable parameters.

Our chosen measure of the quality of the gate, the average fidelity, is just the fidelity of the output of the real gate to the output of the ideal gate, averaged over all possible pure input states. This definition involves integrating over all of the infinite possible input pure qubit states. However, it has been shown that there exists a much simpler, elegant analytic relation, which relates the average fidelity to a simpler quantity, the fidelity of entanglement $F_e$:
\begin{equation}
\bar F=\frac{dF_e+1}{d+1},
\label{eq:fid_ave}
\end{equation}
where $d$ is the dimension of the quantum system (for a two-qubit gate $d=4$) \cite{nielsen2002simple}.

The fidelity of entanglement for a noisy two-qubit gate on a state, $\mathcal{N}_{t_g}[\rho]$, is defined by considering a maximally entangled state of four qubits, with two of the qubits then acted on by the gate. Letting $\ket{\Psi}=\tfrac{1}{2}\sum_{i,j=S,T}\ket{ij,ij}$ be the maximally entangled state of four qubits, with density matrix given by $\rho_{_\Psi}=\ket{\Psi}\bra{\Psi}=\tfrac{1}{4}\sum_{i,j,k,l=S,T}\ket{ij,ij}\bra{kl,kl}$.
We define $\ket{\varphi}$ as the state of the system after it has evolved under the ideal gate, then the fidelity is just $F_e=\bra{\varphi}(\mathcal{N}_{t_g}\otimes{I})[\rho_{_\Psi}]\ket{\varphi}$. Here an ideal version of the gate would act on the state thusly: $\mathcal{N}_{t_g}[\rho]=U\rho U^\dagger$.

The formula for the average gate fidelity~(\ref{eq:fid_ave}) is simple enough that it can be calculated analytically for the gate performed on the dressed qubits in the approximation of equation~(\ref{eq:master_high}).


The first noise regime that we consider is high frequency charge noise, determined by the single qubit dephasing time $T_2$. The evolution of the system subject to high frequency noise only is described by the unshifted master equation in Eq.~\eqref{eq:master} and the reduced form of the polaron shifted master equation Eq.~\eqref{eq:master_high}. We note that in order to derive the analytical approximations for the fidelity we go into an interaction picture with respect to the polaron shifted Hamiltonian, leaving just the second, third and fourth terms in Eq.~\eqref{eq:master_high}, such that in this picture an ideal gate corresponds to the initial state remaining unchanged:
\begin{equation}\begin{split}
\dot\rho_h=\sum_{i=1}^2\gamma_{\phi i}\mathcal{D}[\tilde \sigma_{zi}]\rho_h/2+\Gamma_d\mathcal{D}[(\tilde \sigma_{z1}+\tilde \sigma_{z2})]\rho_h/2.
\label{eq:master_high_bare}
\end{split}
\end{equation}

The second noise regime is low frequency noise, determined by the single-qubit ensemble dephasing time, $T_2^*$. The low frequency noise is modelled by a normally distributed random shift to the qubit splittings, $\delta J_i$. The resulting low frequency qubit noise master equation in the interaction picture with respect to $H_\text{pol}$ is:
\begin{equation}
\dot\rho_\ell=-i[H_n,\rho_\ell]+\Gamma_d\mathcal{D}[(\tilde \sigma_{z1}+\tilde \sigma_{z2})]\rho_\ell/2,
\label{eq:master_low}
\end{equation}
where $H_n=(\delta J_1\sigma_{z1}+\delta J_2\sigma_{z2})/2$. Numerically, we simulate this noise by randomly assigning values $\delta J_i$ in a Gaussian distribution, with the width determined by $T_2^*$, and then averaging over $\delta J_i$.

 For the estimate of the infidelity due to all sources of noise, we again go into the interaction picture with respect to $H_\text{pol}$, and use the complete master equation:
\begin{equation}
\begin{split}
\dot\rho_e=-i[H_n,\rho_e]+\gamma_{\phi1}\mathcal{D}[\tilde \sigma_{z1}]\rho_e/2+\gamma_{\phi2}\mathcal{D}[\tilde \sigma_{z2}]\rho_e/2\\+\Gamma_d\mathcal{D}[(\tilde \sigma_{z1}+\tilde \sigma_{z2})]\rho_e/2\,,
\label{eq:master_int}
\end{split}
\end{equation}
where the first term describes the low frequency noise, the second and third terms describe the qubit dephasing due to high frequency noise and the final term describes the decay of the edge mode.

We can estimate the decay rates associated with both the qubits and the QH edge modes, making use of both data from recent experiments and some theoretical modelling.  For qubit decay rates, we take values from the recent experiment reported in Ref.~\onlinecite{nichol2016high}, where the single qubit decay rate is found to be
$\gamma_\phi=1/T_2=1/(7\text{$\mu$s})$, and the ensemble 
dephasing time, $T_2^*=700\text{ns}$.

The QH edge mode decay rate, $\kappa$, can also be estimated based on recent experiments.  Reference~\onlinecite{mahoney2016chip} observes QH droplets to have loaded quality factors as high as $Q=100$. However, this loaded $Q$ is a pessimistic approximation for an intrinsic $Q$, as it describes a system where the edge modes are coupled to the adjacent metallic contacts.  To estimate the intrinsic $Q$ associated with these devices, we consider several other sources of loss that can affect a QH droplet and explore the limits to which they can be minimised. 

First, we consider what the resistivity of the 2DEG, $\rho_{xx}$, may predict about the intrinsic $Q$.  The model of Ref.~\onlinecite{viola2014hall} for a circular droplet predicts a $Q$ of the fundamental EMP mode of $\rho_{xy}/\rho_{xx}$.  Based on the parameters of the experiment of Reference~\onlinecite{mahoney2016chip}, this effect predicts an intrinsic $Q$ of approximately $1000$.  However, an improved design of the experiment can be expected to reduce $\rho_{xx}$ because, on a quantum Hall plateau, this resistivity is expected to become very low.

Second, we can consider what the effect of coupling between the edge modes and phonons predicts about the intrinsic $Q$.  This effect has been studied in Refs.~\cite{zulicke1997edge,cano2013microwave}.  In the frequency range we are considering, it is predicted that the piezoelectric coupling dominates, in which case the intrinsic $Q$ is independent of the size of the QH disc.  For GaAs, this size-independent quality factor is estimated to be approximately $Q\sim1500$, broadly consistent with the prediction above using resistivity.  For this value of $Q$, for a disc of radius of $R=8\mu$m, this gives an ideal decay rate of approximately $\kappa=17 \times10^6$ rad/s.

\subsection{Fidelity as a function of detuning}

The fidelity of the gate can be calculated by numerically simulating the gate using the master equations of Eqs.~\eqref{eq:master}, \eqref{eq:master_high} and \eqref{eq:master_low}. However, a more intuitive understanding of the sources of error and the physical processes that cause them can be attained by looking at analytic approximations for the fidelity. In this section, we will derive such analytic expressions, as well as describe our numerical simulations of the systems in both high and low frequency noise regimes in order to demonstrate that our approximations are accurate, and to draw conclusions about the dependence of the fidelity on the detuning between the qubit and edge mode frequencies. As a result, we show that the gate fidelity can be maximised by finding the appropriate detuning for the qubit oscillations.

\subsubsection{High-frequency noise}

First, we determine the gate fidelity due to the high-frequency noise of the qubits as a function of detuning. Before deriving analytical expressions for the gate fidelity using only high-frequency noise, we show the equivalence of the master equations in Eqs.~\eqref{eq:master} and \eqref{eq:master_high}, in terms of the resulting qubit evolution. We simulated the master equation~\eqref{eq:master} by expanding harmonic oscillator operators in a Fock state basis. It is essential for the operation of this entangling-gate that photon emission from EMP mode does not occur during the operation of the gate. In the parameter regime of interest, where $\kappa$ might be comparable or significantly larger than $J_{12}$, this requires that the oscillator is not not highly excited during the operation of the gate. Consequently, accurate results are obtained with a moderate dimension size for the Fock space.

\begin{figure}[t]
\centering
\includegraphics[width=0.95\linewidth]{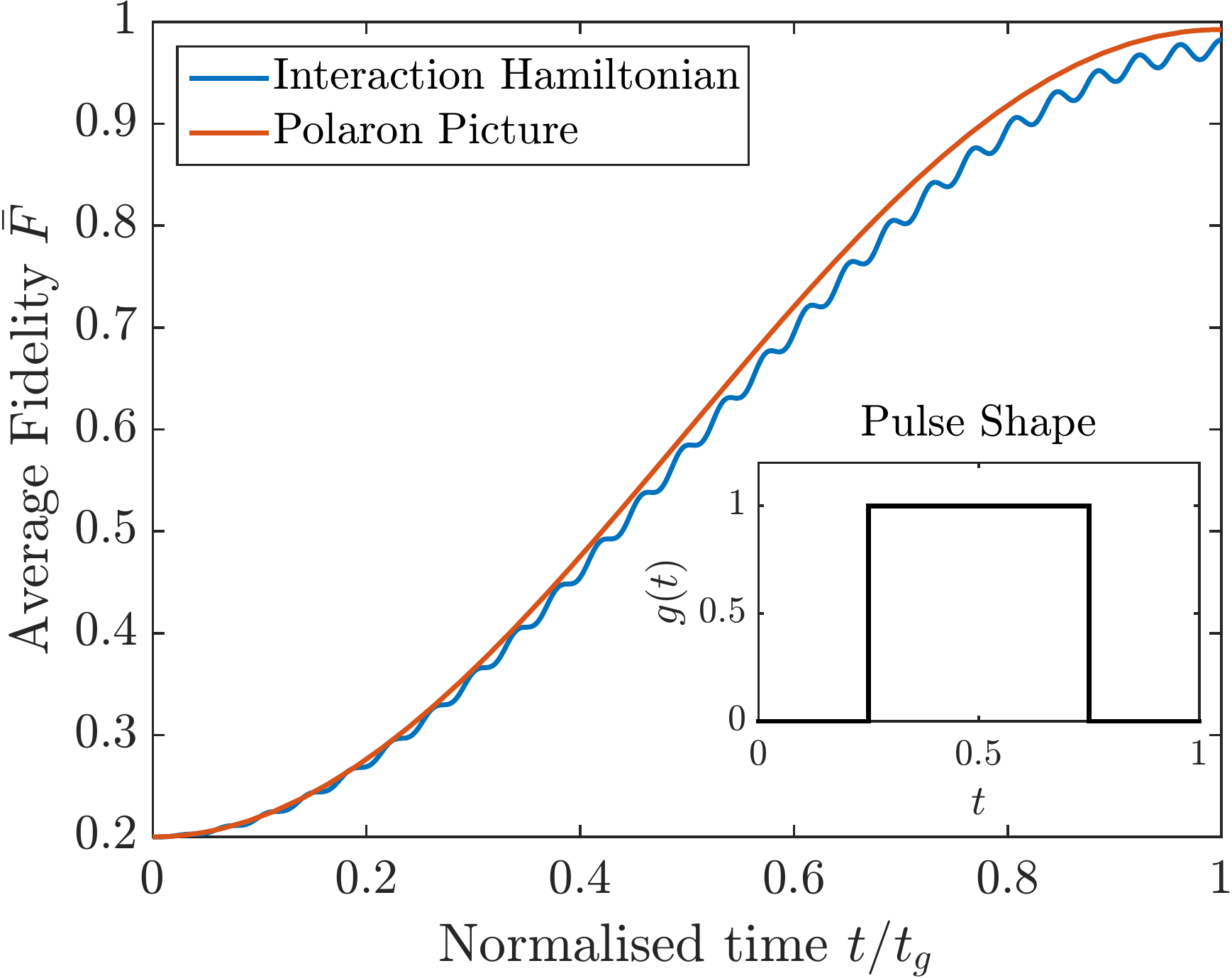}
\includegraphics[width=0.95\linewidth]{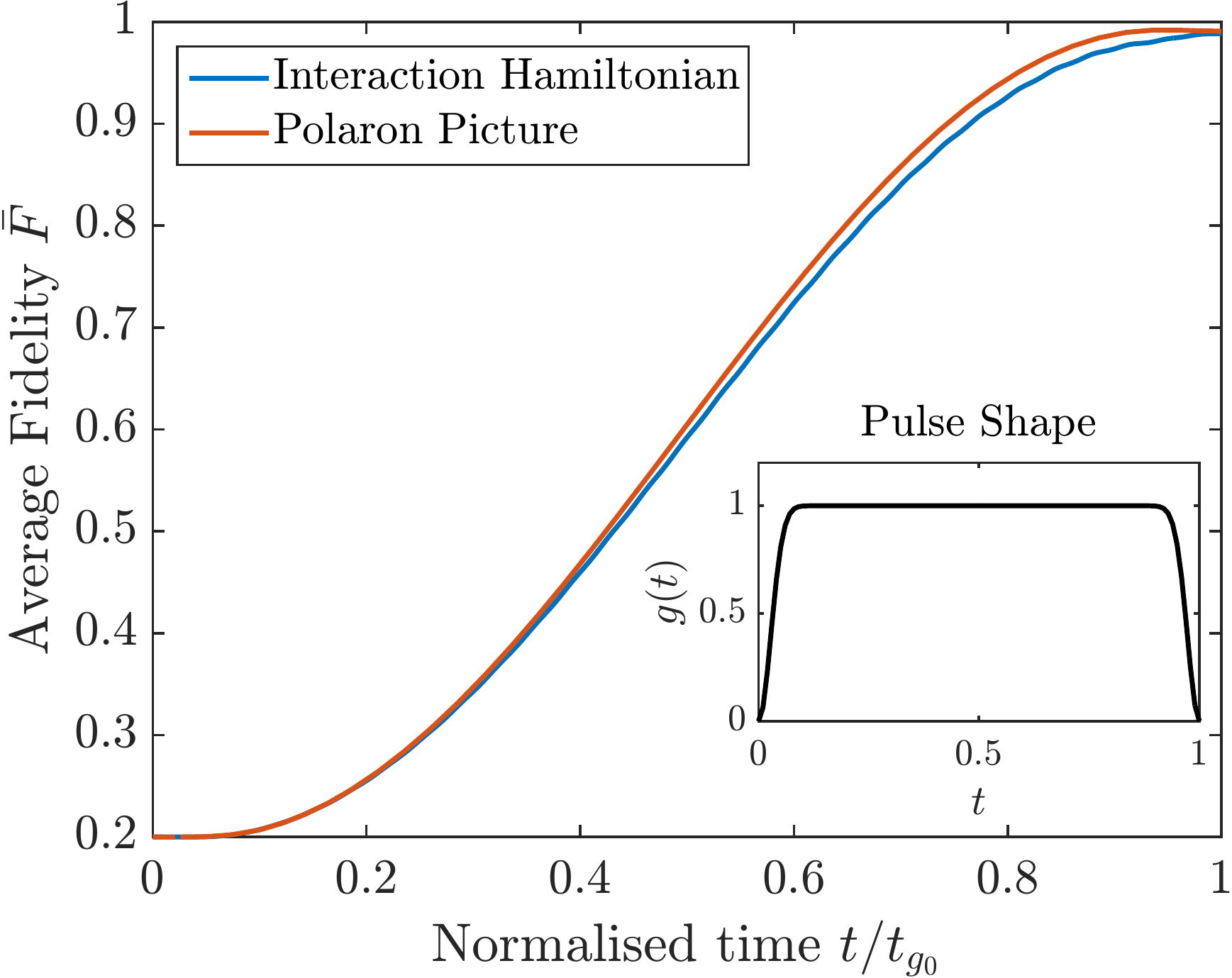}
\caption{Simulation of the gate for a fixed detuning $\Delta/\kappa=300$ using both the polaron shifted (red) and unshifted (blue) descriptions using a square pulse (a) and a shaped pulse (b), with the chosen equation parameters: $g_0=7.29\times10^{8}$ rad/s (see Sec.~\ref{sec:electrostatic_coupling}), $\kappa=17.2\times10^6$ rad/s, and qubit dephasing times of $T_2=7\,\mu$s and $T_2^*=700$ns, from Ref.~\onlinecite{nichol2016high}.  The shaped pulse was chosen to achieve maximum fidelity for this value of detuning.}
\label{fig:square_pulse}
\end{figure}

Fig.~\ref{fig:square_pulse} shows the simulation of the gate in time for both the polaron shifted and unshifted descriptions of the system using a square pulse and a shaped pulse defined by the function:
\begin{equation}
g(t)=g_0(1-\cos^{2n}(t\pi/t_{g_0})), 
\label{eq:pulse_shape}
\end{equation}
where $g_0$ is the electrostatic qubit-edge coupling calculated in Sec.~\ref{sec:electrostatic_coupling}, $t_{g_0}$ is the time necessary to perform the optimum gate when using this shape pulse and $n$ is an integer parametrising the squareness of the shaped pulse. We will adjust the time for the gate $t_{g_0}$  to achieve the highest fidelity possible fidelity for each $n$.
Note that $n=1$ corresponds to a $\sin^2(x)$ pulse that slowly increases to $g_0$; this will result in the adjusted gate time being roughly twice as long as a square pulse: $t_{g_0}= 2t_g$. On the other hand, as $n\rightarrow\infty$, the pulse becomes more square, and the adjusted gate time approaches $t_g$.

As shown in Fig.~\ref{fig:square_pulse}a, when the square pulse is used, the polaron shifted master equation accurately describes the envelope of the evolution of the qubit simulated by the unshifted master equation. However, there are oscillations in the qubit state beneath this envelope, associated with the polarons, which are not included in the results obtained by using the polaron shifted master equation. The envelope of the unshifted master equation also lags behind the polaron picture, resulting in a difference between the fidelity calculated using the shifted and unshifted master equations, $\delta F$, of approximately $\delta F\simeq 0.0045$.

As shown in Fig.~\ref{fig:square_pulse}(b), we can remove such oscillations by slowly turning on and off the coupling, that is, using a pulse described by Eq.~\eqref{eq:pulse_shape} with finite $n$, where in Fig.~\ref{fig:square_pulse}(b) we have used $n=66$. We note that, for the shaped pulse, it is necessary to optimise the adjusted gate time $t_{g_0}$, such that the optimal approximate gate defined by Eq.~\eqref{eq:master} can be performed; for $\Delta/\kappa=300$, the adjusted gate time is $t_{g_0}\simeq1.116t_g$, where $t_g$ is defined in Eq.~\eqref{eq:gate_time}. We note that by carefully tuning the value of our pulse shaping parameter $n$, and the gate time, $t_{g_0}$, we can improve the fidelity of the gate such that, for our chosen parameter of detuning we have an improvement in the average fidelity of $\delta F\approx0.0021$.

We now derive the analytical expression for the high-frequency noise, to which we will compare the numerical simulations from both the polaron shifted and unshifted master equations. As the right hand side of Eq.~\eqref{eq:master_high_bare} is independent of time, the time evolution operator for the noisy gate is proportional to an exponential of a matrix, $\mathcal{N}(t)\propto e^{\lambda_h t}$, where the matrix $\lambda_h$ is determined by Eq.~\eqref{eq:master_high_bare}. We set the two-qubit dephasing rates to be equal ($\gamma_{\phi1}=\gamma_{\phi2}=\gamma_{\phi}$) and use Eq.~\eqref{eq:fid_ave} to derive an expression for the average fidelity for the gate, which only includes the high-frequency noise of the qubit:
\begin{subequations}
\begin{align}
\begin{split}
\bar F_h=& \frac{1}{5}\left(2+2e^{-(\gamma_\phi+\Gamma_d)t_g}+e^{-2\gamma_\phi t_g}/2\right.\\
&\qquad\qquad\qquad\quad\left.+e^{-(2\gamma_\phi+4\Gamma_d)t_g}/2\right)
\label{eq:fid_high_a}
\end{split}
\\
\simeq&\,1-\frac{4}{5}\left(\frac{\pi\gamma_\phi\Delta}{2g^2}\frac{\Delta^2+\kappa^2}{\Delta^2}+\frac{\pi\kappa}{4\Delta}\right),
\label{eq:fid_high}
\end{align}
\end{subequations}
where we have expanded the exponentials to first order, to derive the analytical approximation that can be understood more simply.
The infidelity due to the qubit's high frequency noise has two terms: the first describes the dephasing of the qubit, which is proportional to the detuning, and the second describes the energy loss from the edge modes of the QH disc.

\begin{figure}
\centering
\includegraphics[width=0.95\linewidth]{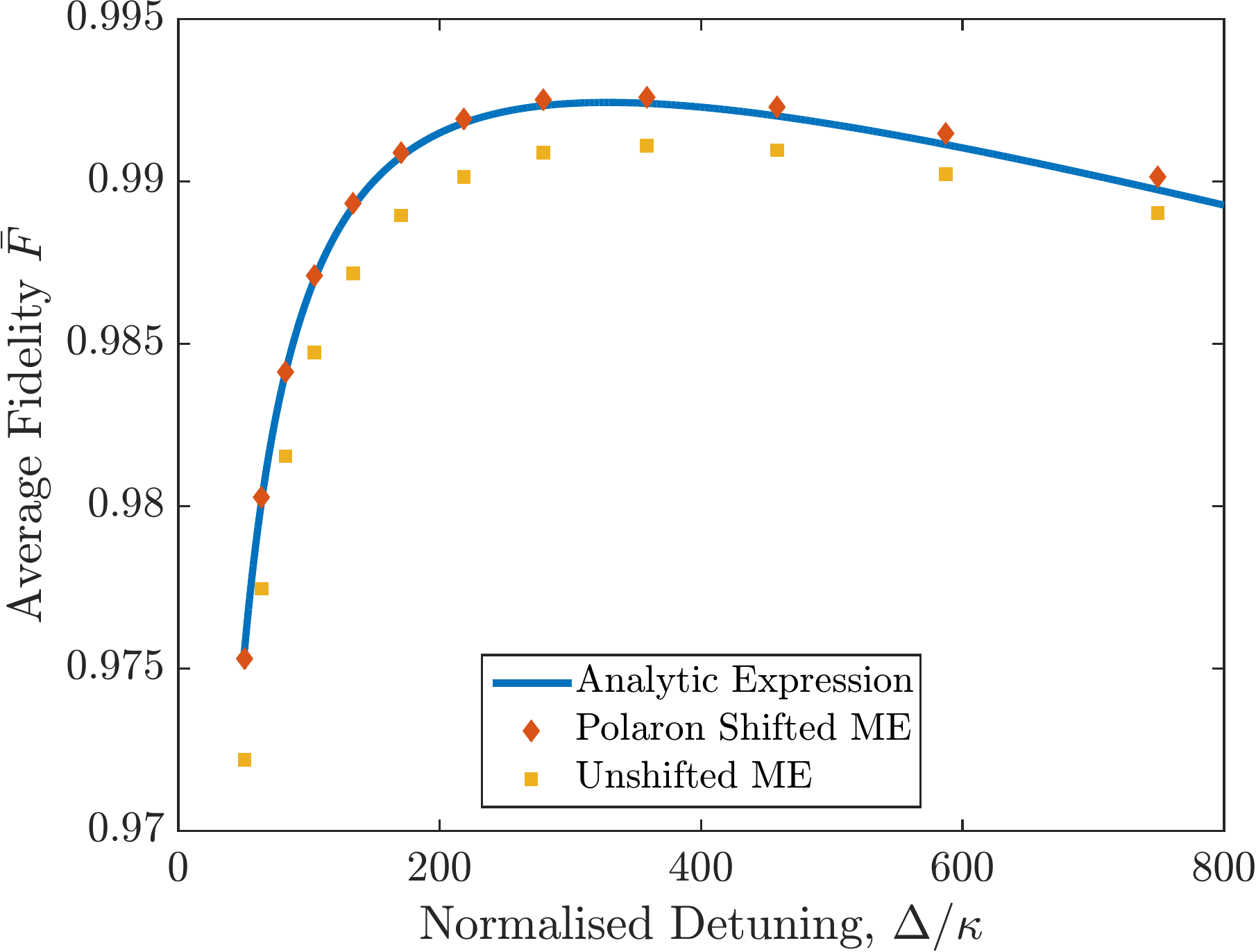}
\caption{Average fidelity as a function of detuning, $\Delta$, for high frequency noise using the analytical expression of Eq.~\eqref{eq:fid_high_a} (solid blue line), and numerical simulations using the polaron shifted master equation of Eq.~\eqref{eq:master_pol} (red diamonds) and and the unshifted master equation from Eq.~\eqref{eq:master} (yellow squares).  Parameters are chosen as in Fig.~\ref{fig:square_pulse}:  $g_0=7.29\times10^{8}$ rad/s, $\kappa=17.2\times10^6$ rad/s, and qubit dephasing time of $T_2=7\,\mu$s.}
\label{fig:fid_delta_high}
\end{figure}

Fig.~\ref{fig:fid_delta_high} presents a comparison between the analytical expression for the average fidelity (making use of the exact exponential expressions from Eq.~\eqref{eq:fid_high_a}) and the full numerical simulations for the average fidelity using the master equations from the polaron shifted approximation in Eq.~\eqref{eq:master_high} and the complete model in Eq. \eqref{eq:master}, as a function of the detuning, $\Delta$, using the same equation parameters as those for Fig.~\ref{fig:square_pulse}.  We note that, for the numerical simulations of the unshifted master equation, we have optimised the adjusted gate time for each value of $n$, and then optimised for each value of $n$ (for $2\leq n\leq 1000$: ranging from very slow to very square) at each data point for the detuning, to achieve the best possible fidelity for any given $\Delta$.  It should be noted that as $\Delta$ increases the optimal $n$ also increases, meaning that a squarer pulse is preferred for larger detuning. As predicted the correction to the fidelity, $\delta F$, decreases as $\Delta$ increases, because to leading order $\delta F$ is some power of $g/\Delta$. Therefore, large detuning results in a closer agreement between the fidelities predicted by $H_\text{pol}$ and $H_\text{int}$.

According to Eq.~\eqref{eq:fid_high}, the error associated with the EMP energy loss can be reduced by going to a very large detuning. However, the gate time is approximately proportional to the detuning, as demonstrated by Eq.~\eqref{eq:gate_time}, so an increase in detuning results in the potential for the qubits to dephase before the gate has been completed. Eq.~\eqref{eq:fid_high} describes a linear relationship between the error caused by high frequency noise dephasing the qubits and the detuning, which results in a turning point in the fidelity as a function of $\Delta$; a turning point clearly visible in Fig.~\ref{fig:fid_delta_high} in all three plots. Therefore, there is an optimal detuning $\Delta_{\text{opt},h}$, demonstrated by the local maxima in Fig.~\ref{fig:fid_delta_high}, which represents a trade off between reducing the noise associated with the qubit dephasing and the noise associated with the EMP losing energy.

There is clearly good agreement between the numerical simulations of polaron shifted master equation of Eq.~\eqref{eq:master_high} and the analytical expression.  We can also see that there is a disparity between the analytical expression and simulation that used the unshifted polaron master equation. This disparity is most pronounced at low detuning, where the distinction between the dressed and bare qubits is most significant, leading to a correction to the fidelity predicted by polaron picture of approximately $\delta F\simeq0.0035$. However, it is also clear in Fig.~\ref{fig:fid_delta_high} that this correction to the fidelity is dependent on $\Delta$, and that as $\Delta$ increases the agreement between the simulation of the unshifted master equation and the analytic expression improves, as predicted in Sec.~\ref{sec:qubit_qubit_coupling}. In principle, the polaron-shifted analytic estimation of the fidelity could be made more accurate for low detuning by performing an expansion in $g/\Delta$. However, this will not affect our analyses or conclusions in any major way, since it is clear that the overall trend of the fidelity as a function of $\Delta$ is qualitatively captured at this level of approximation, and because we are able to simulate the full model.

\begin{figure}
\centering
\includegraphics[width=0.95\linewidth]{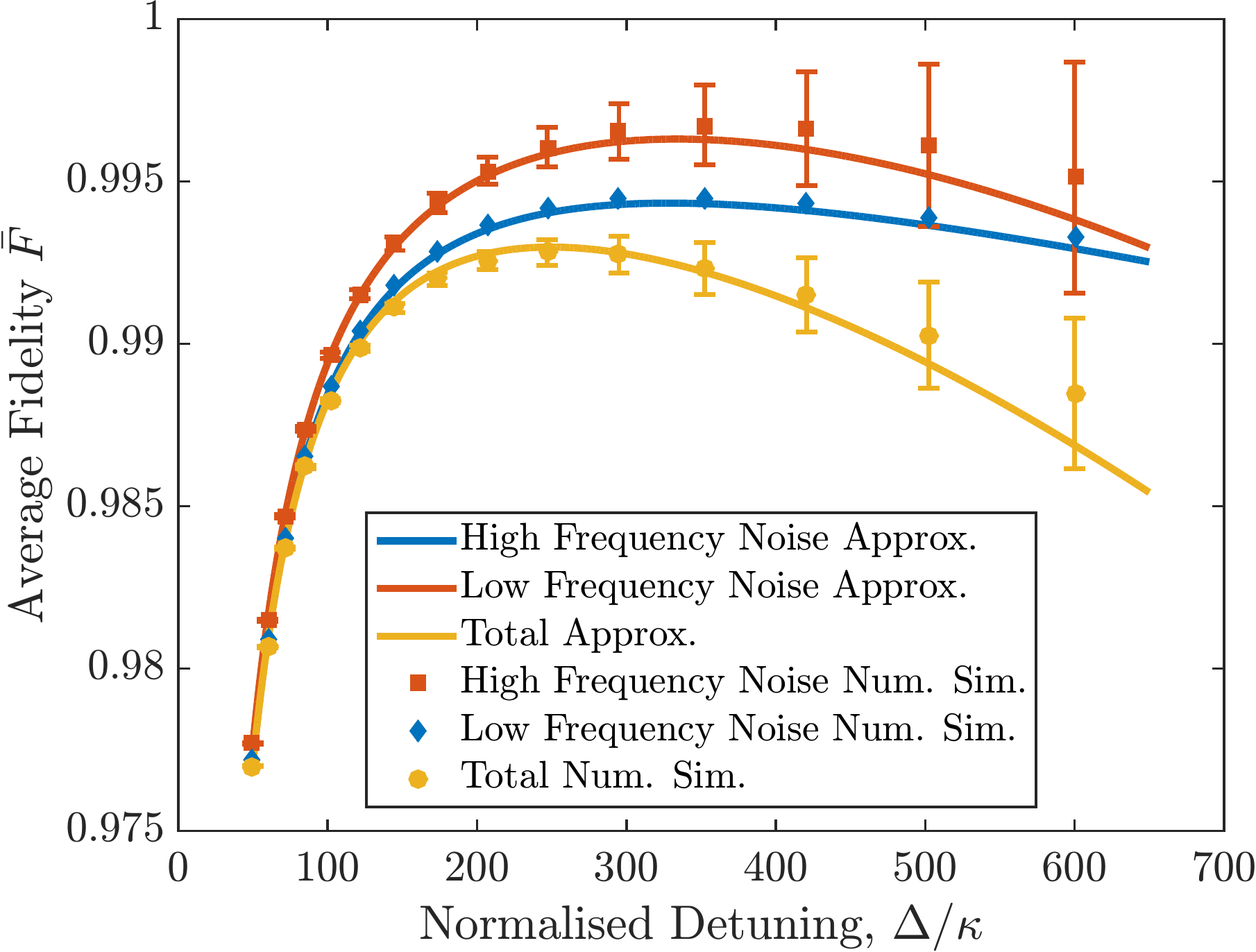}
\caption{Average fidelity as a function of detuning, $\Delta$, for high frequency noise (blue), low frequency noise (red) and for all sources of noise (yellow), using both the analytic expressions (solid lines) of Eqs.~\eqref{eq:fid_high_a}, \eqref{eq:fid_low} and \eqref{eq:fid_ful} and the full numerical simulations in the polaron shifted frame (diamonds, squares, circles).  Parameters are chosen as in Fig.~\ref{fig:square_pulse}:  $g_0=7.29\times10^{8}$ rad/s, $\kappa=17.2\times10^6$ rad/s, and qubit dephasing times $T_2=7\,\mu$s and $T_2^*=700$ ns.}
\label{fig:fid_delta_comp}
\end{figure}

\subsubsection{Low frequency noise}

We now investigate the fidelity in the presence of low frequency noise. 
Again, the right hand side of Eq.~\eqref{eq:master_low} is also independent of time, so solving the master equation results in a time evolution operator for the noisy gate proportional to an exponential matrix, $\mathcal{N}(t)\propto e^{\lambda_\ell t}$.
The first term in Eq.~\eqref{eq:master_low} is imaginary, which results in temporal oscillations in the fidelity, as calculated using Eq.~\eqref{eq:fid_ave}.
The period of the oscillations is determined by the qubit splittings, $\delta J_i$, which can be averaged to give an expression for $\bar{F}$:
\begin{subequations}
\begin{align}
\begin{split}
\bar F_{\ell}=&\frac{1}{5}\left(2+2e^{-(t_g/T_2^*)^2-\Gamma_dt_g}+e^{-2(t_g/T_2^*)^2}/2\right.\\
&\qquad\left.+e^{-2(t_g/T_2^*)^2-4\Gamma_dt_g}/2 \right)
\label{eq:fid_low_a}
\end{split}
\\
\simeq&1-\frac{4}{5}\left[\left(\frac{\pi\Delta}{2T_2^*g^2}\right)^2+\frac{\pi\kappa}{4\Delta}\right].
\label{eq:fid_low}
\end{align}
\end{subequations}
To obtain the second expression, the exponentials have been expanded to first order, giving an analytical approximation for the fidelity due to the qubit's low frequency noise. Again we use the exact expression in Eq.~\eqref{eq:fid_low_a} for graphs and analysis thereof.

Fig.~\ref{fig:fid_delta_comp} presents a comparison between the analytical expressions for the average fidelity and the full numerical simulations for the average fidelity using the polaron shifted master equations \eqref{eq:master_high}, \eqref{eq:master_low} and \eqref{eq:master_int}, so that they can be compared.

Focussing on the low-frequency noise effects, the red curve and points in Fig.~\ref{fig:fid_delta_comp} give a comparison between the analytic expression, as described by Eq.~\eqref{eq:fid_low}, and the full numerical simulations of Eq.~\eqref{eq:master_low}, for the low frequency noise with the ensemble qubit dephasing time $T_2^*=700$ ns (obtained from Ref.~\onlinecite{nichol2016high}). We expect that for low frequency noise contributions to the infidelity, the numerical simulations will be hindered by finite sampling. A finite sample means that very large values of $\delta J_i$ (both positive and negative) are likely to be under-represented, resulting in large uncertainties in the values of fidelity acquired by the simulation, therefore the analytical approximations will capture more of the physics for low frequency noise infidelity.
This effect is shown in Fig.~\ref{fig:fid_delta_comp}, as the error bars grow larger with $\Delta$, which may be understood as there being a larger spread of fidelity values for large values of $\Delta$ due to increased gate times. However, it is clear that there is good agreement between the analytical approximation and the numerical simulations in the regime of detuning close to the optimum.

According to Eq.~\eqref{eq:fid_low}, the infidelity due to the qubit's low frequency noise has two terms: the first describing the dephasing of the qubit, which is quadratic in detuning, and the second describing the energy loss from the edge modes of the QH disc, which is identical to the term seen in Eq.~\eqref{eq:fid_high} for the same error process, as shown in Fig.~\ref{fig:fid_delta_comp}, where the two regimes result in very similar plots, proportional to the reciprocal of the detuning.

Also evident in Fig.~\ref{fig:fid_delta_comp}, Eq.~\eqref{eq:fid_low} describes a turning point in the fidelity as a function of detuning, due to a trade off between minimising the infidelity due to qubit dephasing and EMP energy loss.  Again, as with high frequency noise, there is an optimum detuning $\Delta_{\text{opt},\ell}$. The major difference between the infidelities due to the high and low frequency noises is that, while the infidelity due to the high frequency noise scales linearly with detuning, the infidelity due to the low frequency noise scales quadratically.  This difference is revealed at large values of $\Delta$. 
It is clear then, that near the optimal detuning, the dominant source of noise is high frequency noise. However, it also implies that as we move to very large detuning, $\Delta\gg\Delta_{\text{opt},\ell}$, the low frequency noise overtakes the high frequency noise as the dominant source of error in the gate.

\subsubsection{Full solution with all noise regimes}

Finally, we consider the full master equation of Eq.~\eqref{eq:master_int}. Solving in a similar way to Eqs.~\eqref{eq:fid_high_a} and \eqref{eq:fid_low_a} gives an analytic approximation for the average fidelity including all noise contributions:
\begin{equation}
F_e\simeq1-\frac{4}{5}\left[\left(\frac{\pi\Delta}{2T_2^*g^2}\right)^2+\frac{\pi\gamma_\phi\Delta}{2g^2}\frac{\Delta^2+\kappa^2}{\Delta^2}+\frac{\pi\kappa}{4\Delta}\right].
\label{eq:fid_ful}
\end{equation}
The first term describes the low frequency noise dephasing, the second term describes the qubit dephasing due to high frequency noise, and the final term describes the dephasing of the EMP mode. For the chosen parameters, we have calculated an optimal average gate fidelity of $\bar{F}=0.9930$, with a gate time of $t_g=19.19$ ns. This gate fidelity is a promising result for quantum computation, as threshold gate fidelities have been quoted as low as $F=0.9917$ for performing noisy controlled not gates on surface codes \cite{wang2011surface}, and this estimate is well in excess of the entangling gate fidelities of $F=0.90$ that have been experimentally demonstrated using capacitive coupling techniques \cite{nichol2016high}. As shown by the red plot in Fig.~\ref{fig:fid_delta_comp}, in the limit that $T_2$ can be extended indefinitely, the average gate fidelity reaches $\bar{F}_\ell=0.9972$.

\subsection{Fidelity as function of qubit-edge coupling}

\begin{figure}
\centering
\includegraphics[width=0.95\linewidth]{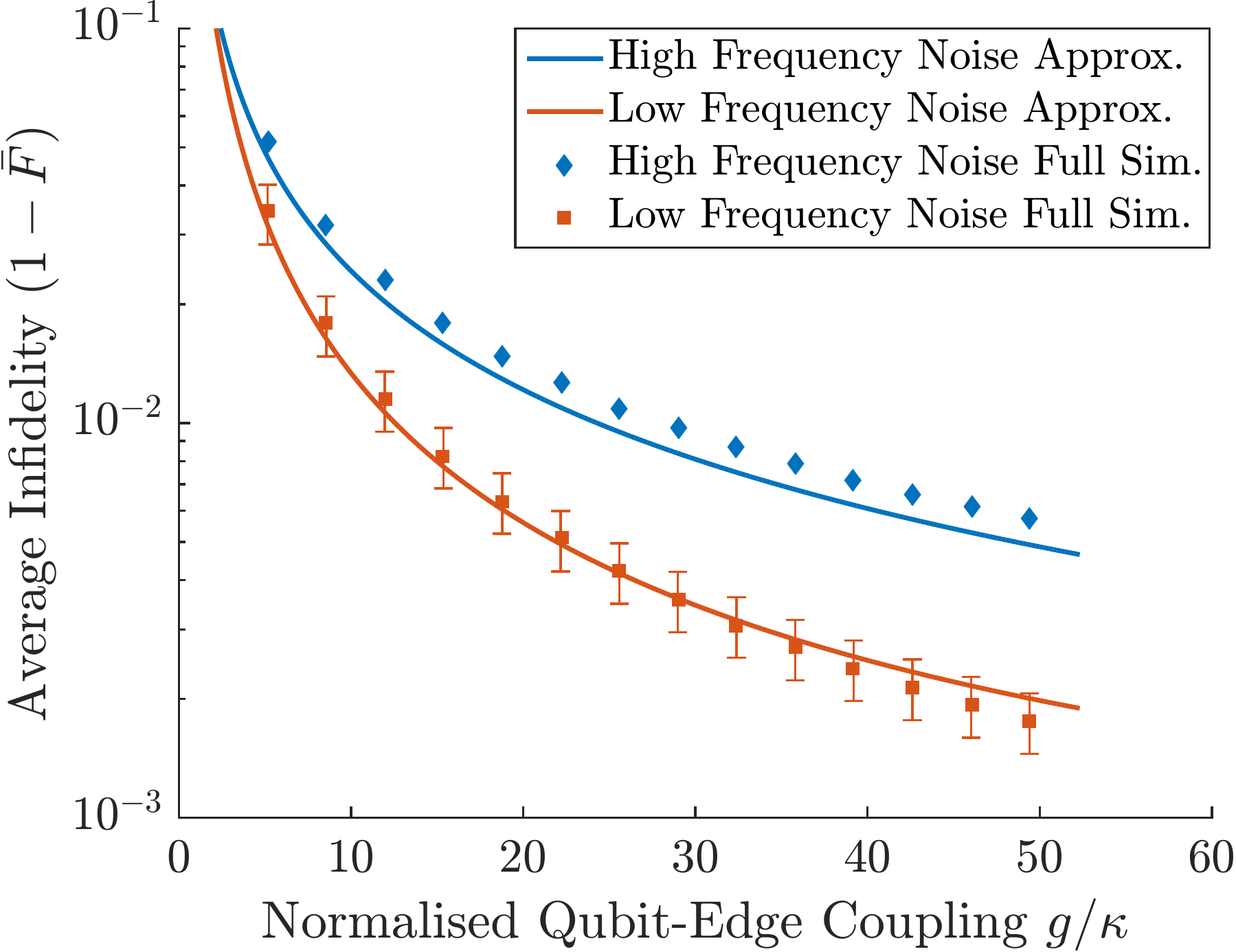}
\caption{Average infidelity as a function of qubit-edge coupling, $g$, at optimal detuning, with the optimal detuning chosen for high frequency noise (blue) and the low frequency noise (red).}
\label{fig:fid_g}
\end{figure}

While these values of average fidelity are very promising, there is potential room for improvement by increasing the qubit-edge coupling $g$.  Considering Eq.~\eqref{eq:coupling_t}, we can see that an increase in $g$ can be achieved by fine tuning a number of parameters: the disc radius $R$; the qubit edge separation $y_1$; the inter-dot separation $\Delta y$; the amount of metal surrounding the system $\eta$; or even the amount of the electron that is shifted into the $\ket{(2,0)}$ state, $\theta_q$. 

We now investigate the expected behaviour of the gate fidelity as a function of $g$.  
Using Eq.~\eqref{eq:fid_high} and \eqref{eq:fid_low} we can find an analytic approximation to the maximum fidelities (for both the high and low frequency sources of noise) as well as the optimal detuning needed to produce these fidelities. 
The resulting optimal fidelity can be expressed in terms of a suitable cooperativity for the system, similar to the result in Ref.~\onlinecite{schuetz2015universal} for example. In the ideal limit that $\kappa\ll\Delta$, we approximate $(\Delta^2+\kappa^2)/{\Delta^2}\sim 1$, leading to a `high frequency noise minimising' optimal detuning
\begin{equation}
\Delta_{\text{opt},h}\simeq g\sqrt{\frac{\kappa}{2\gamma_\phi}}.
\label{eq:opt_hi}
\end{equation}
with a maximum fidelity in terms of the cooperativity, $C_h=g^2/{\kappa\gamma_\phi}$, given by:
\begin{equation}
F_{\text{opt},h}\simeq 1-\pi\frac{4}{5}\sqrt\frac{\kappa\gamma_\phi}{g^2}=1-\pi \frac{4}{5}C_h^{-1/2}.
\label{eq:fid_coop_h}\end{equation}

The `low frequency minimising' optimal detuning noise is:
\begin{equation}
\Delta_{\text{opt},\ell}=\left(\frac{(g^2T_2^*)^2\kappa}{\pi}\right)^{1/3}
\label{eq:opt_lo}
\end{equation}
with a maximum fidelity in terms of cooperativity $C_\ell=(T_2^*g^2/\kappa)$ given by:
\begin{equation}
F_{\text{opt},\ell}\simeq1-\frac{2\pi^{4/3}}{5}\left(\frac{\kappa}{T_2^*g^2}\right)^{2/3}=1-\frac{2\pi^{4/3}}{5}C_\ell^{-2/3}.
\label{eq:fid_coop_l}
\end{equation}

Figure~\ref{fig:fid_g} presents the infidelity of the gate as a function of the qubit-edge coupling ($g/\kappa$), as described by Eqs.~\eqref{eq:fid_coop_h} and \eqref{eq:fid_coop_l}. Clearly visible is the decrease in the gate infidelity as $g/\kappa$ increases, meaning that the qubit-edge coupling $g$ must be at least an order of magnitude larger than the EMP dissipation rate $\kappa$ in order to see gate fidelities greater than $0.9$.  
By considering Eq.~\eqref{eq:coupling_t}, increasing $g$ while maintaining a constant $\kappa$ would require an increase in the inter-dot separation ($\Delta y$), a decrease in the qubit edge separation ($y_1$), or an increase in the portion of the electron that is shifted to the $\ket{S(0,2)}$ state ($\theta_q$), all of which are likely to affect the qubit dephasing rates $T_2$ and $T_2^*$.

\section{Conclusion}

We have proposed and analysed a promising approach for performing long-range high-fidelity entangling gates of singlet-triplet qubits in double quantum dots based on an electrostatic interaction between the charge state of a qubit and the edge modes of a QH droplet.  Based on parameters from recent experiments, we have calculated the electrostatic coupling between a singlet-triplet qubit and the edge of the QH droplet, by considering the difference in energy of the QH edge modes as the qubit is shifted between logical basis states.  By driving oscillations in the state-dependent dipole of the qubit, the effect is a qubit-state-dependent force on the edge mode.  Then, using a polaron transform, we have shown how this interaction mediates a coupling between two qubits, each coupled to the edge of the QH droplet, and that this coupling may be used to implement a two-qubit entangling gate that is dependent on the detuning between the edge mode frequency and the qubit drive frequency.

To investigate the performance of this entangling gate, we have analysed the average gate fidelity in two noise regimes: high frequency noise (associated with the $T_2$ dephasing time) and low frequency noise (associated with the dephasing time $T_2^*$) as a function of the detuning.  For each source of noise, we have identified an optimum detuning for the drive frequency of the qubits in order to maximize the fidelity of the gate.  Based on current experimental values for dephasing times, the fidelity of the gate is predominantly limited by the high frequency noise, with average gate fidelities expected to be $\bar{F}=0.9927$, with a gate time of $t_g=19.19$ ns, but with improved $T_2$ times could reach as high as $\bar{F}=0.9972$.  This fidelity may be further improved by engineering the configuration of the system to increase the qubit-edge coupling, which could perhaps be achieved by producing double quantum dot qubits with larger inter-dot distances or qubits with larger state-dependent dipole moments (increasing the value of $\theta_q$).

The electrostatic scheme proposed here is a step toward the implementation of high-fidelity entangling gates between singlet-triplet qubits in GaAs/AlGaAs heterostructures, but we emphasise that our techniques can be readily adapted to other encodings of qubits that involve a charge degree of freedom, as well as spin qubit implementations in other materials that also support QH liquids. 

\begin{acknowledgments}
We thank Alice Mahoney for detailed comments on the manuscript, and acknowledge discussions with David Reilly, Amir Yacoby, Bert Halperin, Michael Shulman, Thomas Smith and Shannon Harvey.  This work was financially supported by the ARC via the Centre of Excellence in Engineered Quantum Systems (EQuS), project No.~CE110001013. 
\end{acknowledgments}

\end{document}